\documentclass[
aps,%
12pt,%
final,%
notitlepage,%
oneside,%
onecolumn,%
nobibnotes,%
nofootinbib,%
superscriptaddress,%
showpacs,%
centertags]%
{revtex4-1}

\usepackage[utf8]{inputenc} 
\usepackage[T2A,OT1]{fontenc} 
\usepackage[english]{babel} 
\usepackage{amssymb} 
\usepackage{mathtext} 
\usepackage{amsmath} 
\usepackage{mathtools} 
\usepackage{graphicx} 
\usepackage{epsfig}

\usepackage{hyperref}

\begin{document}

\title{A proton-pentaquark mixing and the~intrinsic charm model}
\author{Mikhail Mikhasenko}
\email{mikhail.mikhasenko@ihep.ru}
\affiliation{Institute for High Energy Physics, Protvino, Russia}
\affiliation{Moscow Institute of Physics and Technology, Russia}

\begin{abstract}
A new interpretation of intrinsic charm phenomenon based on the assumption of 
pentaquark $\left| u u d c\bar{c} \right>$ mixing with a proton is offered. 
The structure function of the $c$-quark in the pentaquark is constructed. 
Mixing different states is considered theoretically and using 
experiment data on $D$-meson production and inclusive production of the hidden charm particles.
\end{abstract}

\maketitle

Today there are many articles \cite{Lykasov:2012hf,Freeman:2012ry} related to intrinsic charm concept, which was introduced by Brodsky et al. \cite{Brodsky:1980pb}. In the present article we want to discuss a new interpretation of this phenomenon. As will be shown below, the intrinsic charm problem is closely related to the pentaquark $\left| u u d c\bar{c} \right>$ existence and mixing of that state with the proton.

Authors of ref. \cite{Brodsky:1980pb} made the assumption that Fock state decomposition of the proton wave-function contains a non-negligible $\left| u u d c\bar{c} \right>$ component, which results in the specific distribution of c-quark in proton. According to~\cite{Brodsky:1980pb}, the probability to find $\left| uudc\bar{c} \right>$ configuration, in classical perturbation theory, is given by the expression

\begin{equation} \label{eq:pert}
W(A\to q_1 \dotso q_5) \sim \left| \frac{\left< q_1 \dotso q_5|M|A \right> }{E_A-E_{q_1}-\dotso-E_{q_5}} \right|^2,
\end{equation}
where $q_i$ is the momentum of $i$-th quark, $A$ is an initial state, $M$ is a transition matrix element, $E_i$ is the energy of $i$-th quark.


Transforming the above expression and integrating over all quark's contributions except of one $c$-quark, the authors of \cite{Brodsky:1980pb}, derive the expression the distribution of c-quark:
\begin{equation}
\label{eq:brodsky} P(x) = N x^2 \left[(1-x)(1+10x+x^2) + 6x(1+x)\ln\frac{1}{x} \right],
\end{equation} 
where $N$ is determined by normalization condition for one c-quark $\int P(x) dx = 1$.

The distribution $P(x)$ is shown in fig. 1 by solid line. From eq.\eqref{eq:brodsky} and Fig.1 we can see that average momentum fraction of $c$-quark is equal to $2/7$. As a result, noticeable contribution of intrinsic charm into inclusive production of $D$-, $J/\psi$ or $\chi_c$-mesons should be seen at $\langle x\rangle \approx 2/7$ \cite{Gershtein:1980jb,Litvine:1999sv}.

It is necessary to notice that the probability to find $\left| uudc\bar{c} \right>$ configuration \eqref{eq:pert} was constructed under inherently perturbative assumption, since the factor $1/(E-E')$ is used. Let us consider alternative, essentially nonperturbative, model \cite{Kuti:1971ph}. We construct structure function for $c$-quark in the pentaquark $\left| u u d c\bar{c} \right>$, which has quantum numbers like a proton. The probability to find $i$-th quark with momentum fraction $x_i$ in proton for small $x_i$ can be obtained from Regge asymptotic:
\begin{equation}
 dP_i(x_i) \sim \frac{x_i^{1-\alpha_i}}{\sqrt{x_i^2+\mu^2/P^2}},
\end{equation}
where $\alpha_i$ is the intercept of corresponding leading Regge trajectory. For sea quarks this parameter is equal to 1, in the case of light valence quarks we have $\alpha_{1,2,3}=\alpha_{u,d}(0)\approx 0.5$ from leading Regge $f$, $A_2$ trajectories, while for $c$-quark the intercept of leading $J/\psi$-trajectory is equal to $\alpha_{4,5} = \alpha_c(0) \approx -2.2$ \cite{Kartvelishvili:1977pi,Gershtein:2006ng}. Thus, the structure function of the $n$-particle state has the form
\begin{eqnarray}
 dG(x_1,\dots,x_n) &\sim& \delta\left(1-\sum_{i=1}^n x_i\right) \prod_{i=1}^n \frac{dx_i x_i^{1-\alpha_i}}{\sqrt{x_i^2+\mu^2/P^2}}.
\label{eq:nonpert}
\end{eqnarray}
Integrating this expression in the case of $|uudc\bar c\rangle$ pentaquark over $dx_1\dots dx_3$ we can calculate the structure function of $c$-quark:
\begin{equation} 
G(x) = M x^{-\alpha_5}(1-x)^{-1+\gamma_A+\sum_1^4(1-\alpha_i)}, 
\end{equation} 
where $M$ is a parameter, which stands normalization: $\int G(x) dx = 1$.

In original Kuti-Weisskopf model \cite{Kuti:1971ph} the unknown parameter $\gamma_A~=~3$ determining the sea normalization can be obtain using Drell-Yan-West relation~\cite{Drell:1969km}. Using this value we get 
\begin{equation} 
\label{eq:kuti} G(x) = M x^{2.2}(1-x)^{6.7}.
\end{equation} 
The structure function of the c-quark for the pentaquark $\left| uudc\bar{c} \right>$ \eqref{eq:kuti} is shown in Fig.~1 by dotted line in comparison with the formula~\eqref{eq:brodsky}. 
\begin{figure}[h] 
\centering 
\includegraphics[width=8cm]{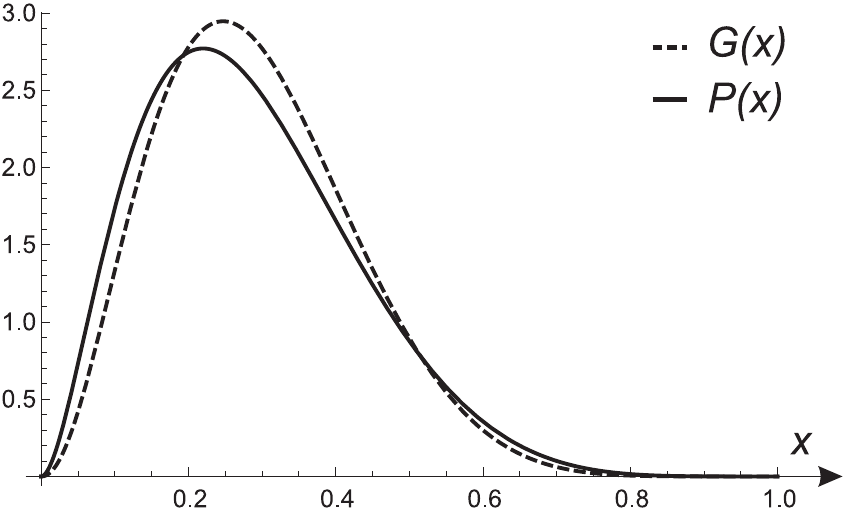} 
\caption{The fraction momentum distribution of $c$-quark in the $\left| u u d c\bar{c} \right>$ state: in the intrinsic charm model \cite{Brodsky:1980pb} satisfied \eqref{eq:brodsky} (solid line);
in the alternative approach based on \cite{Kuti:1971ph} model satisfied \eqref{eq:kuti} (dashed line)} 
\label{fig:str} 
\end{figure} 

As we can see in Fig.~\ref{fig:str} the probability distributions \eqref{eq:brodsky} and \eqref{eq:kuti} are almost the same, under parameters we chose. The result is amazing --- perturbative model \cite{Brodsky:1980pb} and our non-perturbative model, based on Regge trajectories of $c$-quarks give the same probability distribution. The question is appeared, how it could happen. We have one solution only:  the presented in ref. \cite{Brodsky:1980pb} probability distribution refers to the c-quark distribution in the pentaquark $\left| u u d c \bar{c} \right>$ and has nothing to do with sea c-quarks. 

We have one more argument in favor of the hypothesis. It is well known that the quark distribution functions in K-mesons and $\pi$-mesons are different. The difference can be explained by the model~\eqref{eq:nonpert} (see~\cite{Chliapnikov:1977fc}) on the one hand and by~\eqref{eq:pert} model on the other hand. It convinces us that the probability distributions~\eqref{eq:brodsky} and~\eqref{eq:kuti} refer to the valence c-quark momentum distribution in the pentaquark $\left| u u d c \bar{c} \right>$. 

The next problem appeared is the proton-pentaquark mixing which determines absolute normalization. Let matrix element for $p \to \left| uudc\bar{c} \right>$ be equal to $V$, then admixture of the pentaquark in a proton is given by factor $V/(E_1 - E_2)$, where $E_1$, $E_2$ are energies of the pentaquark and the nucleon respectively. The matrix element can be evaluated as $\Lambda_{QCD}$ and the energy difference can be estimated as the difference of masses. In that way the probability to find the pentaquark in a proton is less than: 
\begin{equation} 
W(p \to uudc\bar{c}) \leq \left( \frac{300\,\mathrm{MeV}}{3\,\mathrm{GeV}} \right)^2 \approx 1\% 
\end{equation} 

In fact we should expect the mixing to be much smaller. Most likely it is due to different color state of quarks in the pentaquark and in a proton. There are also experimental constraints on this parameter:
\begin{enumerate} 
\item In the original work \cite{Brodsky:1980pb}, based on assumption that most of the charm cross-section comes from diffraction in pp-interaction, the mixing was evaluated by $1\%$. 
\item In the later work \cite{Hoffmann:1983ah} the restriction $0.59\%$ was given, based on data of European Muon Collaboration (EMC) in hadronic scattering. The best fit to data gave $0.31\%$. 
\item Differential spectrum of $D$-mesons in photo-production processes \cite{Gershtein:1980jb} contradicts the intrinsic charm hypothesis and set the contribution restriction in the range of $\sim 0.1-0.2\%$. 
\item The most severe restriction on the mixing hypothesis can be obtained by hidden charm particles production. Such analysis was performed in ref. \cite{Litvine:1999sv}. According to this paper the upper bound in the $p\to \left| uudc\bar{c} \right>$ probability is $\sim 10^{-7}$.
\end{enumerate} 

The discussed model, based on the assumption about the pentaquark-proton mixing, gives a consistent interpretation of the intrinsic charm phenomenon, which is used for the hadron physics application~\cite{Lykasov:2012hf}. It is worth emphasizing that the observation of the intrinsic charm phenomenon can be result of existence of the stable pentaquark $\left| uudc\bar{c} \right>$. A discovery of pentaquark $\left| uudc\bar{c} \right>$ or getting more stringent restriction is a task of future experiments.

Author would like to thank Prof. A.K.Likhoded and A.V.Luchinsky for productive discussions and a set of constructive remarks. This work was financlally supported by the grant of the president of Russian Federation (\#MK-3513.2012.2).

\bibliography{litr}

\end{document}